# KdV Equation Model in Open Cylindrical Channel under Precession

HAJAR ALSHOUFI [1]

Department of Hydraulic and Water Resources Engineering, Faculty of Civil Engineering, Budapest University of Technology and Economics. Műegyetem rkp. 3., K, 1111 Budapest.

Abstract

A new model for Korteweg and de-Vries equation (KdV) is derived. The system under study is an open channel consisting of two concentric cylinders, rotating about their vertical axis, which is tilted by slope $\tau$ from the inertial vertical $z$, in uniform rate $\Omega_1 = \tau\Omega$, and the whole tank is elevated over other table rotating at rate $\Omega$. Under these conditions, a set of Kelvin waves is formed on the free surface depending on the angle of tilt, characterized by the slope $\tau$, volume of water, and rotation rate. The resonant mode in the system appears in the form of a single Kelvin solitary wave, whose amplitude satisfies the Korteweg-de Vries equation with forced term. The equation was derived following classical perturbation methods, the additional term made the equation a non-integrable one, that cannot be solved without the help of numerical methods. Invoking the simple finite difference scheme method, it was found that the numerical results are in a good agreement with the experiment.

## 1. Introduction.

Most wave phenomena in modern science are decribed by nonlinear equations. The most beautiful model was the one derived in 1895 by two Dutch mathematicians, which the equation takes their names: Korteweg and de-Vries (KdV) equation[2]. Although it was first dedicated for water waves in open channel, after the discovery of solitary wave by Russell (1834), later in the sixties the equation was noticed in other applications like: internal gravity waves in stratified fluid, ion-acoustic waves in plasma, axisymmetric waves in a nonuniformly rotating fluid, nonlinear waves in cold plasmas, axisymmetric magnetohydrodynamic waves and several other physical applications. (Debnath 2012).

In this paper I introduce a new model of this KdV equation with azimuthal dependence where the cylindrical geometry, incorporates both tilt and rotation effects. This is a new case, as all the previous cases that discussed the solitary wave invoked the rotation effect only. The rotation effect, however, is old enough as the internal solitary waves are the famous example that may occur in lakes, oceans, fjords, etc. The earliest experiment was carried out by Maxworthy (1983) to study the rotation effect on the internal solitary wave in a rectangular tank mounted over a rotating table, he noticed that the rotation affected the wave form as it was noticed to be curved backward while moving. He called this wave the internal Kelvin solitary wave, that generated when the pressure gradient normal to the lateral boundary opposes the Coriolis effect generated by the wave motion. The author finally confirmed that the wave speed is independent of the rate at which the system rotates and depends only on the stratification and maximum wave amplitude. A similar experiment was carried out later by Renouard *et al.* (1987) with approximately similar observations of Maxworthy (1983). Their observations about the wave showed that the crest of the wave is neither horizontal nor contained in a vertical plane perpendicular to the side due to rotation effect, but is curved backward, hence there is a spatial phase shift which increases by increasing the distance from the wall, and at a given distance from the wall increases by increasing Coriolis parameter.

---

[1] (E-mail: hajaralshoufi@ymail.com)

[2] The KdV equation was also derived by Boussinesq in the 1870's. He found its first three conservation laws, and its one-soliton and periodic traveling wave solutions, see Equation (30), page 77, of the paper J. Boussinesq, Theorie des ondes et des remous qui se propagent le long d'un canal rectangulaire horizontal, en communiquant au liquide contenu dans ce canal des vitesses sensiblement pareilles de la surface au fond, J. Math. Pures Appl. 17 (2) (1872) 55–108, and Equations (283, 291) of the work J. Boussinesq, Essai sur la theorie des eaux courants, Mem. Acad. Sci. Inst. Nat. France 23 (1) (1877) 1–680. This historical observation is due to Prof. P.J. Olver (~ 2001).

Deep mathematical studies about rotation effect was thoroughly discussed by Grimshaw (1985), who proposed two different models discussing whether the rotation effect is strong or weak. His results showed that if scaling the transverse direction with $\beta$ where $\beta^{-1}$ the dimensionless measure of Rossby radius, if it has either unity value or of order $\varepsilon$ (the shallowness parameter) i.e., strong rotation conditions, then the amplitude of the linear internal Kelvin satisfies a Korteweg-de Vries equation for shallow fluids at the wall, or its counterpart for the deep fluids. The second case, when weak rotation occurs, the scaling of the transverse direction is of order $\varepsilon^2$, then the effects of rotation are small and comparable to those of nonlinearity and dispersion, in this case the transverse variation of wave amplitude is undetermined at the leading order, and is instead contained in the evolution equation, which is a modification of the two-dimensional Korteweg-de Vries equation for shallow fluids, or its counterpart for deep fluids. Although the solitary wave is of the stable type, that does not change its shape while moving, Grimshaw *et al.* (1997) studied the extinction of KdV solution in finite time due to the presence of low-frequency rotational term (additive term to KdV). The extinction time was $2\sqrt{A_0}/\epsilon^2$, where $A_0$ the initial solitary wave amplitude, and $\epsilon = \frac{f^2}{2c}$ (with long wave velocity $c = \sqrt{gh}$, and Coriolis parameter $f = 2\Omega \sin(\varphi)$, $g$ the gravity acceleration, $h$ the depth of the basin, $\Omega$ Earth rotation frequency, and $\varphi$ geographic latitude location) is responsible for low frequency dispersion, which typically describes the effect of rotation in a physical system. They noted that this extinction time could be as important as frictional decay times for internal waves in the ocean.

It was shown that cylindrical symmetric flows can support stationary waves of finite amplitude under certain circumstances. (Leibovich 1970). This encouraged both Derzho and Grimshaw (2002) to study the structure of steady travelling solitary waves of large amplitudes in axially symmetric rotating flows in a long tube including the recirculation zone effect. There was critical amplitude where the flow reversal (recirculation) first occurred. The critical amplitude was related to the incipient rotation and the lowest mode of boundary problem solution at the lowest order. It was noticed that the small parameter that determined the critical width of recirculation area could be related to the amplitude parameter that determined the nonlinearity of the problem, and it took two extreme values, at the minimum level the wave was moving out of the recirculation zone with the normal solitary form, and the recirculation zone reaches a minimum radial extent $\eta_{min}$. And at the maximum level the width of the recirculation zone became infinite, and the crest of the wave looked flattened. In fact, the flow in tubes under rotation conditions was also studied when some obstacle in the direction of the flow exists. Grimshaw (1990) proposed the theory for this case, precisely when upstream flow is nearly resonant (critical) defined with respect to long waves mode, that has a phase celerity almost zero. They used the forced KdV equation for the simulation with forcing term related to the obstacle, and advection term that contained the detuning effect. Later Grimshaw and Zhu (1994) completed the same problem for the unbounded radial swirling flow, they showed that this wave equation is similar to another one derived by Leibovich for freely propagating weakly nonlinear waves on a radially unbounded swirling flow.

The numerical methods that processed Kortweg-de Vries equation under rotation effects are varied, for instance Katsis and Akylas (1987) prepared a numerical scheme of the modified Kadomtsev-Petviashvili (KP) equation to investigate the nature of the waves generated from typical initial conditions. The KP model was solved numerically where the trapezoidal rule was used for performing the integrals, while the equation was discretized through a simple explicit second-order Lax-Wendroff finite-difference scheme. Their results suggested that weakly nonlinear inviscide theory actually revealed the main features of the observed waves, and that wave front curvature is possible because of slow attenuation of the wave amplitude. In their later work similar numerical scheme using Lax-Wendorff finite-difference scheme was used, but this time for the forced KP equation, for the sake of the dependence of soliton radiation on channel width. This equation was derived under conditions of pressure distribution travelling at a speed near the linear-long-wave speed, and it had three-dimensional formulae to express the effects of nonlinearity, linear dispersion, and three-dimensions close to critical speed. The linearized KP equation showed that no steady state is reached at critical conditions. Motivated by studying the dynamics of internal Kelvin waves, Melville *et al.* (1988) studied the effect of nonlinear resonance with Poincaré waves on the stability of weakly nonlinear dispersive Kelvin waves of finite bandwidth, this resonance arose from the nonlinear correction to the speed of the Kelvin wave. Their numerical analysis using Fourier transformation showed that the leading-order Kelvin wave solution in the horizontal direction was unstable, owing to resonant forcing of the transverse velocity associated with the linear Poincaré modes of the channel.

Motivated by studying the nonlinear plasma waves, a group of Chinese researchers Peng *et al.* (2019) carried out the derivation of KP equation in cylindrical geometry, where the effect of ions' densities and their temperature were noticed in the coefficients of the final CKP (cylindrical KP equation). This equation was solved numerically to study three different types of waves: breather, rogue waves, and solitons. The authors invoked the Grammian solutions to solve the soliton problem, and the bilinear Hirota method to solve the breather waves. And finally the rogue equation that was treated in the limit of long wave function that is assumed an exponential polynomial. Their final result showed that the amplitude of rogue waves decreased with the absolute value of time, thus the energy of the wave was damped with the increase of time value. Another novel method used to solve the extended modified KdV equation, using the $\bar{\partial}$-dressing methods, is proposed by Li &Tian (2021). The authors introduced a spectral transform matrix to construct the spatial-time spectral problems of the nonlocal emKdV based on Cauchy-Green integral operator. And then to find the hierarchy of nonlinear evolution equations they used a recursive operator based on the proper spectral transformation matrix, so that the final N-soliton solution is derived and solved.

The new techniques in solving the nonlinear equations are varied, like the inverse scattering transform, the Lax Pairs, the Darboux transformation. However, Hirota bilinear method is of high interest, it was used to solve many type of KdV equation. For instance the form of soliton breaking was solved by Hossen *et al.* (2018), where they discussed the interaction with rogue waves, by introducing different quadratic polynomials that include either *sinh* term or exponential term, and solved the problem accordingly. The (2+1)-dimensional KdV version which has indefinite integrate operator in the direction perpendicular to the dominant one was solved by Wang *et al.* (2019) who proposed a solution based on the Bell's polynomial, where a general bilinear Hirota form was derived. Guo *et al.* (2019) discussed the three dimenional KP equation. Wang *et al.* (2019) studied the dynamics of the generalized normal (3+1) KdV equation. The B-type KP equation, which is weakly dispersive wave propagate in a quasi-media, was also solved based on Hirota bilinear method by Yan *et al.* (2019).

A relatively new case of the solitary wave under rotation conditions was examined by Abderrahmane *et al.* (2011), this wave was observed during the liquid drainage through a centrally located circular opening on the bottom of the container, when shallow water conditions were reached. The free surface first displayed long wavelength increases as the water height decreased, then when the shallow water conditions were reached the free surface undulations transformed almost instantly into a surface bulge or solitary wave rotating around the cylindrical wall of the container. Their case here was treated under swirl conditions using full Euler equations, but by taking the wave variation dominantly along the azimuthal direction of the flow and the radial motion was neglected. In later work by Amaouche *et al.* (2013), the same case study of the wave in the cylindrical tank was treated in different model contained all velocity components plus Coriolis effect. In this case the linearized solution of the boundary value problem leaded to a modified Bessel equation that included rotational effects, and in both cases the material derivative was considered as a function of time and the azimuthal coordinates from which they included the rotational effect of the cylinder itself. Even if the dimensionless scaling was taken into consideration, the normal shallow water speeded up.

In open channels one may find simple linear waves, as well as other nonlinear ones, this includes the weakly and strongly nonlinear waves. Those waves can be generated by navigation, or the operation of hydraulic controls or accidentally, then all waves happened to occur because of them are forced ones, and one the „great primary wave", i.e., the solitary wave, is forced wave that accompanied a canal boat in its motion. (Darrigol 2008). Such phenomena affect the hydraulic conditions in the channel and pose engineering problems. The violent impacts of water waves on walls create velocities and pressures that are bigger in magnitude than the normal ones, which may increase the erosive action and warrant the reinforcement or redesign of the channel. Also, one should not forget the dynamics of sediment transport processes generated by the action of the flow phenomena. Thus, this work in the present paper is motivated by studying the forced oscillations and their instabilities in open channels. Precisely, studying the solitary wave which is carried out here theoretically by deriving the new KdV version, and experimentally by comparing the mathematical model with the experiment. This wave is also called the single Kelvin fundamental resonance mode. As the interested reader can see the previous work (**cfr**. ALSHOUFI 2021) about the forced oscillations that were studied linearly first, it turned out that when the sytem resonates the wave becomes singularly large and the linear analysis is not valid anymore. Thus, to complete the linear analysis in this work I introduce the weakly nonlinear case where this specific wave is under consideration, because it has similar features to the famous solitary wave first observed by Russell (1843). The paper is divided as follows: in **§2** The governing equations where the new system of coordinates after transformations from the laboratory frame to frame fixed on the tank is introduced, in addition to flow assumptions, in **§3** experiment and results where a description of the design is carried out with comparing the results experimentally, in **§4** conclusion.

## 2. Governing Equations.

The motion of the wave is supposed to occupy in cylindrical annulus that is confined between the inner cylinder with radius $r_{min}$, and the outer cylinder with radius $r_{max}$. This distance does not exceed 10cm, and it was deduced by the work of Akylas and Katsis (1987) that in such small geometries if the width $2b$ smaller than specific ratio which is $b \leq \bar{h}^2/A$, where $\bar{h}$ the average depth of water, $A$ the typical wave amplitude, the three-dimensional effect can be negligible. This ratio was computed in the designed flume for relatively big amounts of water inside it, those are the cases where the solitary wave was observed. Table (1) presents those results. Thus, the motion is assumd two-dimensional one, where the predominant direction is the azimuthal one, as water was the main liquid fluid used in the experiment, then the flow is assumed incompressible and inviscid one. Additional assumption was on replacing velocity components (which are $u, w$ in the azimuthal and axial directions, respectively) with scalar potential function $\phi$, and assume the flow irrotational one. As the surface of the wave is under the atmospheric conditions then the pressure is assumed negligible. The flume was filled with different volumes of water that are characterized with their average depth $\bar{h}$. The evolution of the free surface function is a time-space depenent function $\eta(\theta, t)$. Any solitary wave appear on the surface has an amplitude $A$. In this work we assume that the total mechanical energy of the flow is conserved in the fully developed flow, thus, turbulence, wall friction due to boundary layers, internal friction due to viscosity, capillarity due to the surface tension and dissipation of vorticity are all connected to energy dissipation and will be neglected. As mentioned that the channel is tilted from the axial direction, this is characterized by the slope of the upper tilted table $\tau$. (**cfr**.§3 Experimental Procedure). As the solitary wave is a long wave, thus the platform in studying this wave is the shallow water theory, this theory depends on the hydrostatic pressure distribution, and one important feature of this theory is that no assumption is made regarding the magnitude of surface elevation or the velocity components. (Stoker1957). Thus, the parameter that describes the evolution of waves in shallow water is called the shallowness parameter and it is going to appear in the system of equation when introducing the scaling process for the dimensionless analysis. This is also called the dispersion parameter $\sigma$:

$$\sigma = \frac{\bar{h}}{\lambda} \ll 1.$$

where $\lambda = 2\pi r_{max}$ represents the length scale or the horizontal scale, which is assumed to be the periphery of the outer cylinder. In the experiment this ratio was all the time satisfied, as it is clear in Table (1).

Table (1). Two-dimensional and shallowness effects in the channel.

| Volume (ml) | $A$(m) | $\bar{h}$(m) | $b$(m) | $\bar{h}^2/A$(m) | $\tau$ | $\sigma$ |
|---|---|---|---|---|---|---|
| 10000 | 0.112 | 0.0933 | 0.049 | 0.078 | 0.0167 | 0.0666 |
| 12000 | 0.0886 | 0.112 | 0.049 | 0.142 | 0.0117 | 0.0799 |
| 14000 | 0.1024 | 0.1307 | 0.049 | 0.167 | 0.0233 | 0.0933 |

As usual z-axis is taken vertically upward, the bottom is given by: $z = 0$, while the free surface as: $z = \bar{h} + \eta(\theta, t)$, the flow is confined between the inner and the outer radius of the channel. The boundary conditions on rigid walls like the bottom satisfy the impermeability condition so that the normal velocity is zero: $w = 0$ at $z = 0$. At the free surface of water, the kinematic condition is given as:

$$w = \frac{\partial \eta}{\partial t} + \frac{u}{r}\frac{\partial \eta}{\partial \theta}, \tag{1}$$

We can now write Euler equations as the following:

$$\frac{Du}{Dt} = g\tau\sin(\Omega t - \theta) - \frac{P_\theta}{r.\rho} - 2\Omega\tau\cos(\Omega t - \theta)w + \Omega^2\tau\sin(\Omega t - \theta)z - O(\tau^2), \tag{2}$$

$$\frac{Dw}{Dt} = -g - \frac{P_z}{\rho} + 2\Omega\tau\cos(\Omega t - \theta)u - \Omega^2\tau\cos(\Omega t - \theta)r - O(\tau^2), \tag{3}$$

where $\frac{D}{Dt}$ represents the material derivative operator:

$$\frac{D}{Dt} = \frac{d}{dt} + \frac{u}{r}\frac{d}{d\theta} + w\frac{d}{dz}. \tag{4}$$

The conservation of mass or continuity equation can be written as:

$$u_\theta + rw_z = 0, \tag{5}$$

The condition for irrotational flow is worth to be split as follows:

$$\xi_r = \left(\frac{1}{r}\frac{\partial w}{\partial \theta} - \frac{\partial u}{\partial z}\right) = 0, \tag{6}$$

The treaties provided by Korteweg-de-Vries (1895) showed that to find the nonlinear type of wavy motion represented by KdV equation, there should be other important variable to measure the nonlinearity is called the amplitude parameter, and it is going to be included in the process of dimensionless analysis as well, it has the form:

$$\varepsilon = \frac{A}{\bar{h}}.$$

Thus, the unknown quantities of velocity field components, the pressure and water surface are first going to be scaled as follows to include the shallowness effect:

$$u = c\bar{u}, w = \sigma c\bar{w}, \bar{\theta} = \theta, \tau = \sigma\bar{\tau}, \bar{z} = \frac{z}{H}, \bar{r} = \frac{r}{\lambda}, \bar{\eta} = \frac{\eta}{H}, t = \frac{\lambda}{c}\bar{t}, P = g\rho(H - z) + \rho g H \bar{P}, \Omega = \frac{c}{\lambda}\bar{\Omega},$$

Where the long wave speed is given by: $c = \sqrt{g\bar{h}}$. Where we assume that $H = \bar{h}$. After substitution into the major equations of continuity, momentum, and bounday conditions, we find:

$$\overline{u_\theta} + \overline{rw_z} = 0, \tag{7}$$

$$\frac{D\bar{u}}{D\bar{t}} = \bar{\tau}\sin(\bar{\Omega}\bar{t} - \bar{\theta}) - \frac{\bar{P}_\theta}{\bar{r}} + \sigma^2[-2\bar{\Omega}\bar{\tau}\cos(\bar{\Omega}\bar{t} - \bar{\theta})\bar{w} + \bar{\Omega}^2\bar{\tau}\sin(\bar{\Omega}\bar{t} - \bar{\theta})\bar{z}], \tag{8}$$

$$\sigma^2\frac{D\bar{w}}{D\bar{t}} = -\bar{P}_z + \sigma^2[2\bar{\Omega}\bar{\tau}\cos(\bar{\Omega}\bar{t} - \bar{\theta})\bar{u} - \bar{\Omega}^2\bar{\tau}\cos(\bar{\Omega}\bar{t} - \bar{\theta})\bar{r}], \tag{9}$$

$$\bar{w} = \frac{\partial \bar{\eta}}{\partial \bar{t}} + \frac{\bar{u}}{\bar{r}}\frac{\partial \bar{\eta}}{\partial \bar{\theta}} \text{ on } \bar{z} = 1 + \bar{\eta}, \tag{10}$$

with the irrotational condition:
$$\frac{\sigma^2 \overline{w_\theta}}{\bar{r}} = \overline{u_{\bar{z}}}.$$

In effect equations (7), (8), (9), (10) represent the shallow water equations at the leading order of $O(\sigma)$, as it is obvious from (9) that this scaling leads to the hydrostatic pressure distribution, the main assumption of shallow water theory. One can also notice at the leading order that the effect of both Coriolis and Euler forces in equations (8) and (9) are of second order of the shallowness parameter $O(\sigma^2)$, thus, they vanish and only pressure and gravity act. The KdV derivation can be further pursued depending on equations (7), (8), (9), and (10) by taking higher orders of the factor $\sigma$ this was done by Keller (1947) for instance: it leads to fully nonlinear solution. However, we can instead make it easier, especially since our equations are more complicated and here comes the usage of the amplitude parameter $\varepsilon$:
$$(\bar{u}, \bar{w}, \bar{p}, \bar{\tau}, \bar{\eta}) \to \varepsilon(\bar{u}, \bar{w}, \bar{p}, \bar{\tau}, \bar{\eta}),$$

We find after rescaling the following:
$$\overline{u_\theta} + \overline{r w_{\bar{z}}} = 0, \tag{11}$$

$$\bar{u}_{\bar{t}} + \varepsilon\left(\frac{\bar{u}\bar{u}_{\bar{\theta}}}{\bar{r}} + \bar{w}\bar{u}_{\bar{z}}\right) = \bar{\tau}\sin(\bar{\Omega}\bar{t} - \bar{\theta}) - \frac{\bar{P}_{\bar{\theta}}}{\bar{r}} + \sigma^2[-\varepsilon 2\bar{\Omega}\bar{\tau}\cos(\bar{\Omega}\bar{t} - \bar{\theta})\bar{w} + \bar{\Omega}^2\bar{\tau}\sin(\bar{\Omega}\bar{t} - \bar{\theta})\bar{z}], \tag{12}$$

$$\sigma^2\left[\bar{w}_{\bar{t}} + \varepsilon\left(\bar{v}\frac{\bar{u}\bar{w}_{\bar{\theta}}}{\bar{r}} + \bar{w}\bar{w}_{\bar{z}}\right)\right] = -\bar{P}_{\bar{z}} + \sigma^2[\varepsilon 2\bar{\Omega}\bar{\tau}\cos(\bar{\Omega}\bar{t} - \bar{\theta})\bar{u} - \bar{\Omega}^2\bar{\tau}\cos(\bar{\Omega}\bar{t} - \bar{\theta})\bar{r}], \tag{13}$$

$$\bar{w} = \frac{\partial \bar{\eta}}{\partial \bar{t}} + \varepsilon\left(\frac{\bar{u}}{\bar{r}}\frac{\partial \bar{\eta}}{\partial \bar{\theta}}\right), \tag{14}$$

Equation (14) is the free surface kinematic condition that is applied at: $\bar{z} = 1 + \varepsilon\bar{\eta}$. KdV equation has beautiful balance between quadratic nonlinear term and the linear dispersive one (Grimshaw 2007), this can be invoked by assuming:
$$\varepsilon = O(\sigma^2),$$
Or equivalently for simplicity:
$$\sigma^2 = \varepsilon.$$
So, we find after substitution:
$$\overline{u_\theta} + \overline{r w_{\bar{z}}} = 0, \tag{15}$$

$$\bar{u}_{\bar{t}} + \varepsilon\left(\frac{\bar{u}\bar{u}_{\bar{\theta}}}{\bar{r}} + \bar{w}\bar{u}_{\bar{z}}\right) = \bar{\tau}\sin(\bar{\Omega}\bar{t} - \bar{\theta}) - \frac{\bar{P}_{\bar{\theta}}}{\bar{r}} + \varepsilon[-\varepsilon 2\bar{\Omega}\bar{\tau}\cos(\bar{\Omega}\bar{t} - \bar{\theta})\bar{w} + \bar{\Omega}^2\bar{\tau}\sin(\bar{\Omega}\bar{t} - \bar{\theta})\bar{z}], \tag{16}$$

$$\varepsilon\left[\bar{w}_{\bar{t}} + \varepsilon\left(\frac{\bar{u}\bar{w}_{\bar{\theta}}}{\bar{r}} + \bar{w}\bar{w}_{\bar{z}}\right)\right] = -\bar{P}_{\bar{z}} + \varepsilon[\varepsilon 2\bar{\Omega}\bar{\tau}\cos(\bar{\Omega}\bar{t} - \bar{\theta})\bar{u} - \bar{\Omega}^2\bar{\tau}\cos(\bar{\Omega}\bar{t} - \bar{\theta})\bar{r}], \tag{17}$$

$$\bar{w} = \frac{\partial \bar{\eta}}{\partial \bar{t}} + \varepsilon\left(\frac{\bar{u}}{\bar{r}}\frac{\partial \bar{\eta}}{\partial \bar{\theta}}\right) \text{ at } \bar{z} = 1 + \varepsilon\bar{\eta}, \tag{18}$$

$$\bar{w} = 0 \text{ at } \bar{z} = 0,$$

The irrotational condition then becomes:
$$\varepsilon \frac{\overline{w_\theta}}{\bar{r}} = \overline{u_{\bar{z}}}.$$

Before expanding in terms of the amplitude parameter for the unknown quantities, it is better to integrate the equation replacing the nonlinear terms with the corresponding irrotational versions, and also by using the scalar velocity potential function for each velocity projection: $\bar{u} = \frac{\phi_\theta}{r}$, $\varepsilon\bar{w} = \phi_{\bar{z}}$, so that we get the Forced Bernoulli Equation in the system, where the forcing term is the integral of azimuthal gravity force projection radially as in (19):

$$\phi_{\bar{t}} + \varepsilon\left(\frac{\phi_\theta^2}{2r^2}\right) + \frac{\phi_{\bar{z}}^2}{2} + \bar{\eta} - r\bar{\tau}\cos(\bar{\Omega}\bar{t} - \bar{\theta}) = 0. \tag{19}$$

Now the expansion of the unknown quantities (the amplitude and velocity potential) will be in terms of the amplitude parameter and takes the form:
$$\phi = \phi^{(0)} + \varepsilon \phi^{(1)} + \varepsilon^2 \phi^{(2)} + \cdots,$$
$$\bar{\eta} = \bar{\eta}^{(0)} + \varepsilon \bar{\eta}^{(1)} + \varepsilon^2 \bar{\eta}^{(2)} + \cdots.$$

Let's re-write the equations, mainly the continuity, Bernoulli, the kinematic and boundary conditions, respectively. We get:

$$\varepsilon \left[\frac{\phi_{\bar{\theta}\bar{\theta}}}{\bar{r}^2}\right] + \phi_{\bar{z}\bar{z}} = 0 \quad \text{at} \quad 0 < \bar{z} \leq 1 + \varepsilon \bar{\eta},$$

$$\phi_{\bar{t}} + \varepsilon \left(\frac{\phi_{\bar{\theta}}^2}{2\bar{r}^2}\right) + \frac{\phi_{\bar{z}}^2}{2} + \bar{\eta} - \bar{r}\bar{\tau}\cos(\bar{\Omega}\bar{t} - \bar{\theta}) = 0,$$

$$\phi_{\bar{z}} = \varepsilon \left(\bar{\eta}_t + \varepsilon \left(\frac{\phi_{\bar{\theta}}}{\bar{r}^2} \bar{\eta}_\theta\right)\right) \quad \text{at} \quad \bar{z} = 1 + \varepsilon \bar{\eta},$$

$$\phi_{\bar{z}} = 0 \quad \text{at } z = 0.$$

The unified form of the phase of tilt $\Omega t$, with the azimuthal angle $\theta$ in the forced term can be harnessed in favor of d'Alambert wave solution, as in the first order of the problem we have:

$$\phi_{1\bar{z}} = \bar{\eta}_{0\bar{t}} = -\phi_{0\bar{t}\bar{t}},$$
$$\phi_{1\bar{z}} = -\frac{z}{\bar{r}^2}\phi_{0\bar{\theta}\bar{\theta}},$$

This will give the linear long wave equation:

$$\phi_{0\bar{t}\bar{t}} - \frac{1}{\bar{r}^2}\phi_{0\bar{\theta}\bar{\theta}} = 0.$$

which has d'Alembert solution in the dimensional form as:

$$\phi(\theta, t) = f(r\theta - ct) + g(r\theta + ct), \qquad (20)$$

where $c$ here is assumed the mean azimuthal velocity that the wave in the channel flume is assumed to rotate with about the outer periphery, thus it has the form:

$$c = r\Omega.$$

Thus, in the dimensionless form of the problem we can write:

$$\phi(\theta, t) = f(\theta - \Omega t) + g(\theta + \Omega t), \qquad (21)$$

Hence, we can introduce the following:

$$\bar{\theta} - \bar{\Omega}\bar{t} = \vartheta,$$
$$\partial \bar{\theta} = \partial \vartheta,$$
$$\partial \bar{t} = -\Omega \partial \vartheta,$$

We also assume slow time evolution for the wave by introducing:

$$T = \varepsilon \bar{t}.$$

Substituting back into Bernoulli and continuity, and dropping the bars we get:

$$\varepsilon \left[\frac{1}{r^2}\phi_{\vartheta\vartheta}\right] + \phi_{zz} = 0, \qquad (22)$$

$$-\Omega \phi_\vartheta + \varepsilon \left(\phi_T + \frac{\phi_\vartheta^2}{2r^2}\right) + \frac{\phi_z^2}{2} + \eta - r\tau\cos(\vartheta) = 0, \qquad (23)$$

$$\phi_z = \varepsilon \left(-\Omega \eta_\vartheta + \varepsilon \left(\eta_T + \frac{\phi_\vartheta}{r^2}\eta_\vartheta\right)\right). \qquad (24)$$

Equations (22), (23), (24) are going to be expanded in terms of $\varepsilon$. It is worth to mention up to this point that the tilt angles in this system are in general small ones, and that the cases where the solitary wave were noticed corresponded to very small tilt character factor: $\tau$, (cf. Table 1), thus it is assumed that if $\varepsilon \to 0$ (which is the solution we are interested in for small wave amplitude) then $\tau \to 0$, so that at the leading order of the problem the forced term can be neglected.

$$\phi_{0zz} = 0 \quad \text{in} \quad 0 < z < 1,$$
$$\phi_{0z} = 0 \quad \text{at} \quad z = 1,$$
$$\phi_{0z} = 0 \quad \text{at} \quad z = 0,$$
$$-\Omega\phi_{0\vartheta} + \frac{\phi_{0z}^2}{2} + \eta_0 = 0.$$

The most important result at the leading order then is the fact that the connection between potential velocity and the amplitude is the dimensionless rotation speed:

$$\eta_0 = \Omega\phi_{0\vartheta}. \tag{25}$$

Collecting similar terms in higher orders will lead to the final version of KdV equation in the channel under precession conditions, is given as:

$$\left(\frac{\Omega}{2r^2} - \frac{1}{6\Omega r^4}\right)\eta_{\vartheta\vartheta\vartheta} + \frac{3}{\Omega r^2}\eta\eta_\vartheta + 2\eta_T + r\Omega\tau\sin(\vartheta) = 0. \tag{26}$$

This equation is KdV type which unfortunately does not have integrable solution and the usage of numerical method is carried out in the next section. Of course, the form in (26) is dimensionless and using the assumed scaling the dimensional version is recovered to be compared with the experiment. The final version in the dimensional form takes the form:

$$A\eta_{\theta\theta\theta} + B\eta\eta_\theta + C\eta_t + D = 0,$$

where the coefficients A, B, C, D, are respectively given by:

$$A = \frac{1}{\varepsilon H}\left(\frac{\frac{\lambda}{c}\Omega}{2\left(\frac{r}{\lambda}\right)^2} - \frac{1}{6\Omega\frac{\lambda}{c}\left(\frac{r}{\lambda}\right)^4}\right), \quad B = \frac{\frac{3}{(\varepsilon H)^2}}{\frac{\lambda}{c}\Omega\left(\frac{r}{\lambda}\right)^2},$$

$$C = \frac{2\lambda}{\varepsilon^2 cH}, \quad D = \frac{\tau\Omega r}{\varepsilon\delta c}\sin(\theta - \Omega t).$$

If we assume periodic initial condition for $\eta$ as:

$$\eta \approx r\Omega\tau\cos(\theta - \Omega t).$$

then the forced term can be approximately assumed to be:

$$-r\Omega\tau\sin(\theta - \Omega t) \approx \eta_\theta.$$

Thus, equation (26) if it is rescaled into the dimensional form can be written as:

$$A\eta_{\theta\theta\theta} + B\eta\eta_\theta + C\eta_t + \eta_\theta = 0.$$

Then the linearization of this equation by searching solution of the form solution: $\eta = e^{i(k\theta - \Omega t)}$, will lead to the dispersion relation between the frequency and the wavenumber takes the form:

$$\Omega = \frac{k - Ak^3}{C}.$$

- **Finite Difference Scheme**

The finite difference method is the oldest and the simplest numerical method. The modern version is the spectral method, which is not going to be considered here. The KdV equation has three different differential terms, the third derivative, and the quadratic nonlinear term, and the variation with time of the problem. Each term can be extracted using Taylor series expansion. Starting from the third derivative, we are interested in the central difference, so we can write:

$$\eta_{i-2} = \eta_i - 2\Delta\theta\frac{\partial\eta}{\partial\theta} + 4\frac{\Delta\theta^2}{2}\frac{\partial^2\eta}{\partial\theta^2} - \frac{8\Delta\theta^3}{6}\frac{\partial^3\eta}{\partial\theta^3}, \tag{27}$$

$$\eta_{i+2} = \eta_i + 2\Delta\theta\frac{\partial\eta}{\partial\theta} + 4\frac{\Delta\theta^2}{2}\frac{\partial^2\eta}{\partial\theta^2} + \frac{8\Delta\theta^3}{6}\frac{\partial^3\eta}{\partial\theta^3}, \tag{28}$$

$$\eta_{i-1} = \eta_i - \Delta\theta\frac{\partial\eta}{\partial\theta} + \frac{\Delta\theta^2}{2}\frac{\partial^2\eta}{\partial\theta^2} - \frac{\Delta\theta^3}{6}\frac{\partial^3\eta}{\partial\theta^3}, \tag{29}$$

$$\eta_{i+1} = \eta_i + \Delta\theta\frac{\partial\eta}{\partial\theta} + \frac{\Delta\theta^2}{2}\frac{\partial^2\eta}{\partial\theta^2} + \frac{\Delta\theta^3}{6}\frac{\partial^3\eta}{\partial\theta^3}, \tag{30}$$

By adding (27), (28) and then (29), (30), we get:

$$\eta_{i+2} + \eta_{i-2} = 2\eta_i + 4\Delta\theta^2 \frac{\partial^2 \eta}{\partial \theta^2},$$

$$\eta_{i+1} + \eta_{i-1} = 2\eta_i + \Delta\theta^2 \frac{\partial^2 \eta}{\partial \theta^2}.$$

The same when subtracting (27), (28) and then (29), (30) we get:

$$\eta_{i+2} - \eta_{i-2} = 4\Delta\theta \frac{\partial \eta_i}{\partial \theta} + 8\frac{\Delta\theta^3}{3} \frac{\partial^3 \eta}{\partial \theta^3},$$

$$\eta_{i+1} - \eta_{i-1} = 2\Delta\theta \frac{\partial \eta_i}{\partial \theta} + \frac{\Delta\theta^3}{3} \frac{\partial^3 \eta}{\partial \theta^3}.$$

By manipulating and re-arranging the previous equations we get:

$$\frac{\eta_{i+2} - \eta_{i-2}}{4\Delta\theta} - \frac{\eta_{i+1} - \eta_{i-1}}{2\Delta\theta} = \left(\frac{2\Delta\theta^2}{3} - \frac{\Delta\theta^2}{6}\right) \frac{\partial^3 \eta}{\partial \theta^3},$$

Which finally leads to:

$$\frac{\partial^3 \eta}{\partial \theta^3} = \frac{\eta_{i+2} - 2\eta_{i+1} + 2\eta_{i-1} - \eta_{i-2}}{2\Delta\theta^3}. \tag{31}$$

The first order differentiation also is going to be treated using central difference which is:

$$\frac{\partial \eta}{\partial \theta} = \frac{\eta_{i+1} - \eta_{i-1}}{2\Delta\theta}, \tag{32}$$

And finally forward Euler finite difference for the time is assumed as follows:

$$\frac{\partial \eta}{\partial t} = \frac{\eta_i^{n+1} - \eta_i^n}{\Delta t}, \tag{33}$$

Then the final equation is given as:

$$\eta_i^{n+1} = \eta_i^n - \Delta t \frac{A}{C} \frac{\eta_{i+2} - 2\eta_{i+1} + 2\eta_{i-1} - \eta_{i-2}}{2\Delta\theta^3} - \Delta t \frac{B}{C} \eta_i \frac{\eta_{i+1} - \eta_{i-1}}{2\Delta\theta} - \frac{\Delta t D}{C}. \tag{34}$$

- **Implementation**

The grid in the azimuthal direction is going to be discretized depending on the number of points extracted from the $X_{mm}$ grid, which is different for each case depending on the extracted pixels and their corresponding millimeters, where the $mm$ grid is built on the outer periphery of the cylinder and has center at the paper clip number 5 as it is clear in Figure (2) that is why we have a domain of negative and positive values. Those $X_{mm}$ points are relatively big like for the case of 10000ml we have $m = 749$ points, for the case of 12000ml we have $m = 899$ points, and for the case of 14000ml we have $m = 649$ points, the starting point of the domain is the first value of each angle and it is called $a_1$, and the ending point is $b_1$, the initial guess of the problem is assumed a Gaussian fit, where it takes its values based on the extracted pixles from which we determine both the average $avg$ and the standard deviation $st$. The distance between each point of the pixel coordinates causes instability problem using Courant number defined in (35) below, unless small value is introduced to ensure equation (35) is satisfied, which is a specific ratio between the time and the distance represented by Courant number $C_r$:

$$C_r = \frac{\Delta t (u+c)}{r_{max} \Delta\theta}, \tag{35}$$

Where $u$ is assumed to be the mean velocity or the stream velocity, and $c$ is the disturbance speed. It is well known that choosing the value of Courant number bigger than one will certainly lead to instability of the assumed solution, that is why it is preferable to take value between zero and one. In the experiment, the time difference was chosen so that we satisfy the following:

$$\Delta t = C_r \frac{\Delta\theta \, r_{max}}{(\Omega r_{max} + \sqrt{gH})}. \tag{36}$$

## 3. Experiment and Results.

Full details of the experimental apparatus and procedure are in (ALSHOUFI 2021). In order to track the different oscillations in the channel under precession conditions, a real flume was constructed for this purpose. The Figure (1) shows the model, that consists of two concentric cylinders elevated over two different tables, the lower one rotates uniformly about the vertical axis, and the second one rotates and tilts about the axial axis of the cylinders. The tilt of the upper table is adjustable through three different screws distributed in 120° apart from each other, the center of this table is mounted on a support column through a Cardano type universal ball joint. The joint contains two mutually perpendicular horizontal axles around which the upper table can freely tilt in any direction, but it prevents any rotation around the vertical axis. The tilt is being represented by the slope of the upper table and is characterized by $\tau$. This value varied during the experiments, it was noticed that the localized wave that has the solitary wave characteristics can only appear in the channel for small slopes, the results discussed below include relatively big volumes of water like 10000ml, 12000ml, 14000ml. Table (2) shows some characteristic and geometrical features of the design. This includes the three adjustable control parameters of the system: **(i)** The first is the amount of water in the tank, which can be quantified either by its volume or, preferably, by the mean water depth $H = \bar{h}$, by which we ensure the mean water level constraint, which is one-one relationship between volume of water and the average water level:

$$V = \int_0^{2\pi} \int_{r_{min}}^{r_{max}} \int_{z_0}^{z_0+h_{ref}+\eta(t,\theta,r)} r\,dz\,dr\,d\theta = \bar{h}(r_{max}^2 - r_{min}^2)\pi. \qquad (37)$$

**(ii)** The second is the tilt angle, characterized by its tangent $\tau$ as mentioned above, which is also the largest slope of the tilting table. **(iii)** The third is the angular velocity $\Omega$ of the rotating table. The energy is provided to the channel by using DC electrical motor that has external electrical supply, which is battery. The whole system is placed on the laboratory floor by a metal base, noting that currently the rotating table can be turned only in the counterclockwise sense. The inner cylinder has wall length higher than the outer one, the inner one was covered from inside with white papers A4 in size to reduce the light reflection. The motion inside the channel was tracked using JAI PULNiX TM-1405 GE camera, which has Charge-Coupled Device (CCD) sensor type. It has high picture resolution $1392 \times 1040$, with focal length 8mm, and pixel resolution $4.65 \times 4.65 \mu m$.

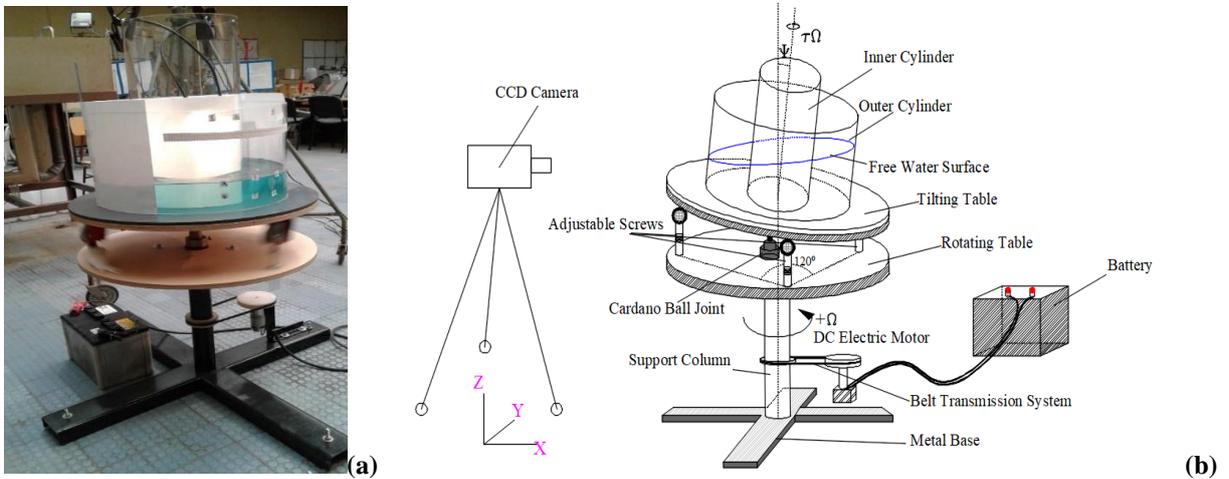

**Figure (1). (a).** The Real Flume Channel in the Laboratory. **(b).** Sketch of experimental setup. The camera system in front of the flume almost at the same level of the water level inside the tank.

**Table (2).** Geometrical information of the channel.

| Notation | Value | Description |
|---|---|---|
| $R = r_{max}$ | 223 mm | Outer Radius |
| $\beta R = r_{min}$ | 125 mm | Inner Radius |
| $2b$ | 98 mm | Channel Width |
| $\beta = \dfrac{r_{min}}{r_{max}}$ | 0.5605 | Radial Ratio |
| h | 260 mm | Maximum Wall Height |
| $A_r$ | 0.1071 m² | Base Area |
| $\forall$ | 0.0275 m³ | Maximum Volume |
| $\tau$ | 0.0117 … 0.115 | Tilt |
| $H$ | 20… 140 mm | Mean Water Level |
| $\Omega$ | 1.5… 8 rad/sec | Angular Velocity |
| $z_0$ | $\approx 50\ mm$ | Elevation of Flume Bottom |

The pictures of this camera were extracted first, and then they were calibrated using a program was designed by the current author for this purpose, it depended on Zhang's method (2000), and the algorithm was applied is the one introduced by Burger (2016), with slight modification that included more radial distortion coefficients. The camera was set at fixed position in front of the channel where it could take continuous pictures, the frequency of capturing the pictures was 1 frame per 30 second. The outer cylinder was also covered with some white papers so that only the place in front of the camera was not covered for better vision. The outer illumination for the system was two bulbs fixed on the inner cylinder's wall to cover the whole inside area, and other external big lamp which was adjustable in its height was sometimes used and fixed in the laboratory floor next to the camera.

The waves that may form in such system are called the inertial forced oscillations, and the linear problem of their solutions leads to a set of modes that are called the Kelvin modes. Such oscillations appear as long as their frequencies smaller than twice the natural one, thus in this system, they all have the same frequency, but they may differ in their wavelength. The solitary wave in the Figure (2) is a solitary Kelvin wave, which accords with the fundamental mode $n = 1$ in the axial direction of motion, $m = 1$ radially, and azimuthal base mode $\rho_1 = 2$. This result was extracted when analyzing the linear case of this problem, from which it was found that resonance in the system occurs when the wavelength matches the outer periphery of the channel. (**cfr.** ALSHOUFI 2021).

The observations of single Kelvin mode in the channel was first for the smallest tilt $\tau = 0.0117$, for different volumes of water varied from 2000ml to 12000ml, the crest was little flattened, and had symmetric form, stable, and of course rotating without changing i its speed, approximately like the wave in the Figure (2), but it looked as if it is going to break, the crest slightly proceded the whole wave body. During the computation of those waves for relatively small volumes 2000ml, to 6000ml, it was noticed that the nonlinearity is much bigger than the dispersion. This leads to assume that such Kelvin solitary mode can be considered for bore development (EL 2007), and that work also is being carried out by the present author. Increasing the volume of water in the channel, leads to bigger amplitudes thus bigger wave lengths, like the wave in the figure (3) for instance. Those cases look really like solitary waves, symmetric in form, not changing their speeds as propagating ahead, but unfortunately they are not stable, they preserve this smooth symmetric form for only one round about the outer periphery of the channel, then disperse into waves have smaller amplitudes and flattened crests. In connections with the forced oscillations in such systems under precession, similar observations in closed tanks were first noticed by McEwan (1970). He called this phenomenon *resonance collapse,* where big waves retain their smooth form for relatively short time, then they collapse. Where the original form cannot be regained even if the flow waxes and wans. And in fact, the computations for the wave in Figure (3) showed that slight priority for dispersion at the expense of nonlinearity, still in good balance to be computed using KdV equation form assumed in (26), using its numerical estimation in (34). Taking time history for the wave in such case is presented in Figure (5), and Figure (4) shows how the wave evolves with time, which shows how flattened the crest with time becomes.

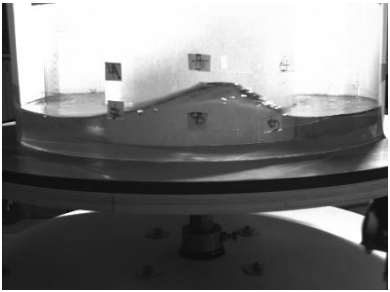

**Figure (2).** Kelvin solitary wave, volume 4000ml, $\tau = 0.0117$, $\Omega = 4.24$rad/sec.

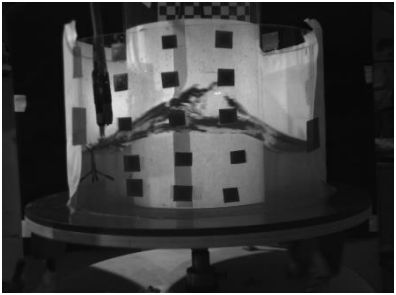

**Figure (3).** Kelvin solitary wave, volume 10000ml, $\tau = 0.0167$, $\Omega = 6.84$rad/sec

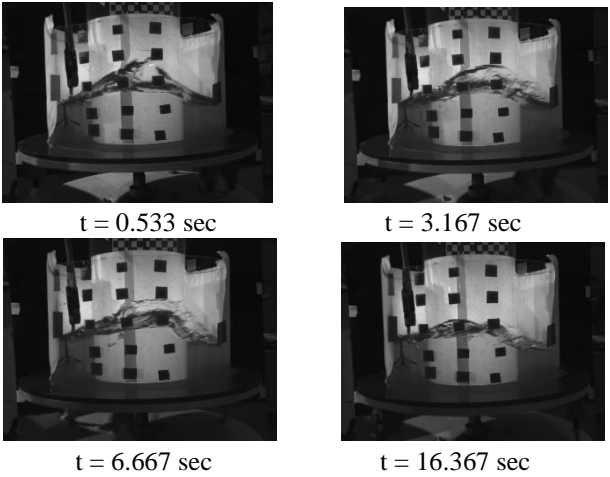

t = 0.533 sec       t = 3.167 sec

t = 6.667 sec       t = 16.367 sec

**Figure (4).** The real wave variation with time.

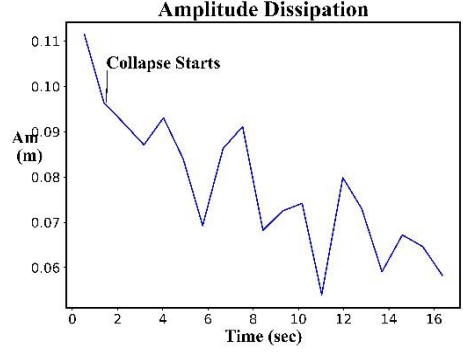

**Figure (5). (a).** Amplitude history of single wave's peak oscillation $\Omega = 6.84$ rad/s, volume 10000ml, mode $\tilde{\varphi}_{1m2}$, $\tau = 0.01667$.

The finite scheme model introduced in the previous section in equation (34) is applied to the observed waves in different volumes, and comparison has been carried out. The initial condition was a Gaussian initial fit from the experimental data:

$$\eta_0 = A_m \cdot e^{-\left(\frac{\theta - avg}{st}\right)^2}.$$

Where $(avg)$ the average of azimuthal data points, $(st)$ their standard deviation, $A_m$ the wave amplitude. For instance in Figure (6) the results of the wave in Figure (3) is presented, where $\Delta\theta = 0.001778$, $\Delta t = 1.243 \times 10^{-6}$, $C_r = 0.01$, volume 10000ml, $\tau = 0.0167$, $\Omega = 6.84$rad/s. All results were captured at a specific time and the solution corresponds to this time only. In Figure (7) other wave was noticed with volume 12000ml, by assuming $\Delta\theta = 0.001594$, we get $\Delta t = 1.25 \times 10^{-6}$, $C_r = 0.01$, $\tau = 0.0167$, $\Omega = 6.018$rad/s. Another one is presented in Figure (8), for the case of volume 14000ml, $\Omega = 6.822$rad/sec, $\tau = 0.0233$, $\Delta\theta = 0.002243$, $\Delta t = 1.66 \times 10^{-6}$, but this wave is almost in breaking form, the crest proceeds the whole body.

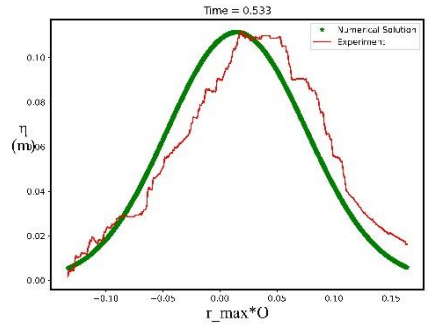

**Figure (6).** Finite difference solution derived in (34), $\bar{h} = 0.0933$, $\tau = 0.01667$, $\Omega = 6.84$rad/s thick line for numerical solution, and the curvy one from the experiment.

As no exact analytical solution for the derived equation, an analytical solution for the problem was derived as proposed by Brauer (2000) with little change as in equation (38). It is kind of fit solution for the results, and it was compared with the experiments at each time, as it is clear in Figure (10).

$$\eta(t,\theta) = A_m sech^2\left(\sqrt{\frac{B\Omega r_{max}}{3C}}\left(\frac{\theta}{2} - \frac{c*t}{2C}\right)\right). \tag{38}$$

To track the wave evolution with time we construct the solution on the outer periphery of the cylinder to show the far field, Figure (9) shows how the wave evolves with time, the equation is not stable thus the need for weight coefficients is harnessed to keep it stable with time, the nonlinear term is multiplied with 0.01, and the dispersion term was multiplied with 0.001, the number of points $m = 70$, and the time difference 0.00176. Of course, the wave starts to disperse under rotation effect, and the system is not stable at all, but the procedure of wave evolution with time shows that the wave starts to decay to the right which is similar to the real decay depicted in Figure (4). However, we cannot assume according to Figure (4) that we still have the solitary wave in essence, but this is the sample of how this system works.

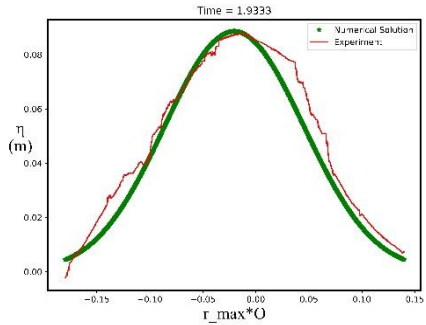
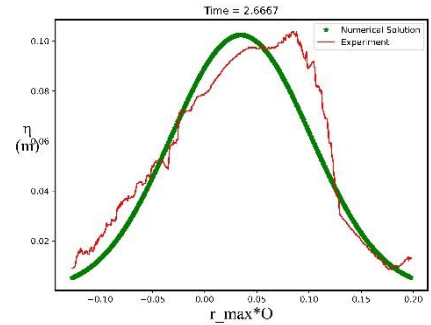

**Figure (7)**. Finite difference solution derived in (34) $\bar{h} = 0.112$m, $\Omega = 6.018$ rad/sec. $\tau = 0.0117$

**Figure (8)**. Finite difference solution derived in (34) $\bar{h} = 0.1307$m, $\Omega = 6.822$ rad/sec. $\tau = 0.0233$.

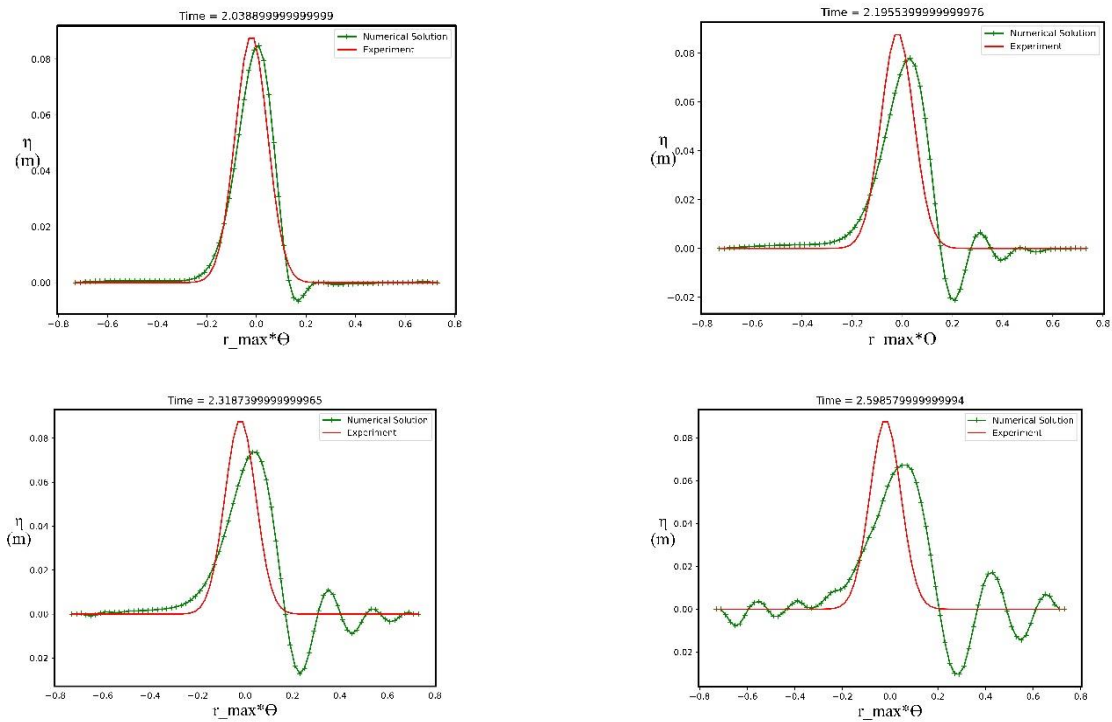

**Figure (9)**. Wave evolution with time, volume 12000ml, $\tau = 0.0117$, the smooth curve is experiment Gaussian intial guess, and the dotted one is the numerical solution.

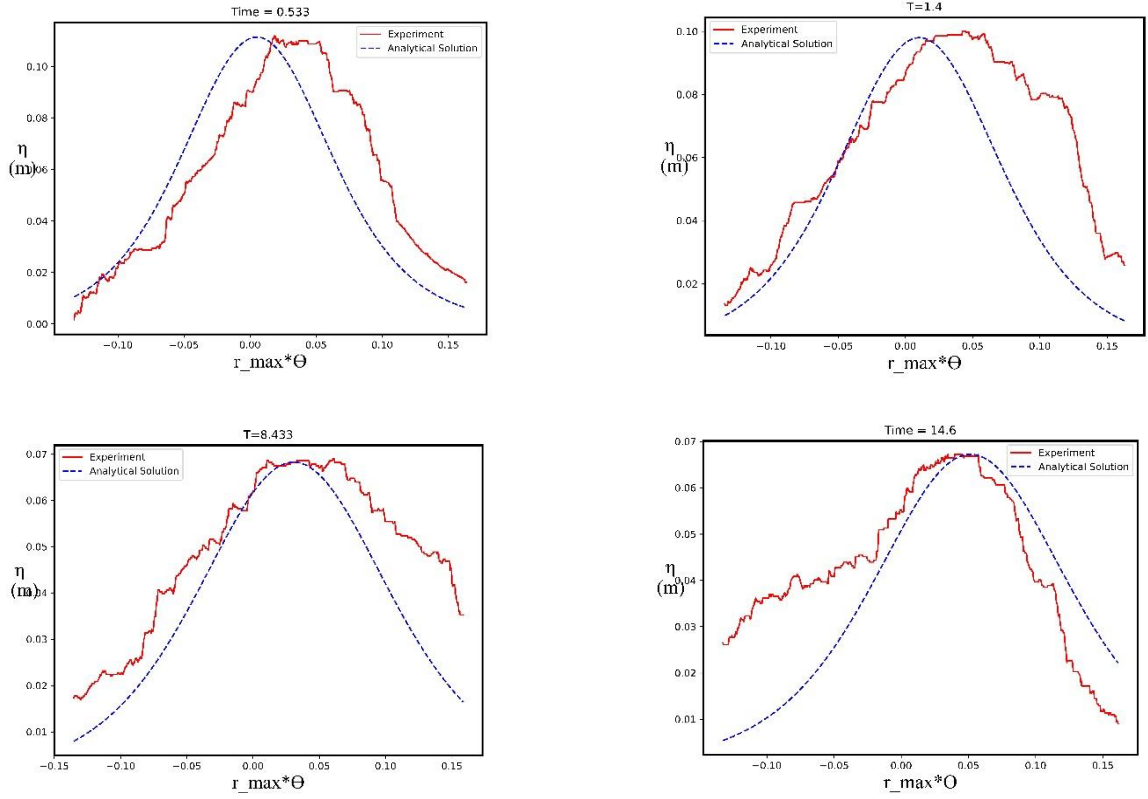

**Figure (10).** Analytical solution derived in (38) with time $\bar{h} = 0.0933$m, $\tau = 0.01667$, $\Omega = 6.84$rad/s.

The solitary wave really can be formed in such systems if the amount of water is big enough, but this wave also can suffer from breaking. It is well known in natural cases, out of rotational effects, from the gas dynamic theory that waves which travel in one direction only, their crests travel faster than troughs, so that the front of each crest continuously steepens, until it breaks down. (Ursell 1953). In Figure (11) we got such collapse of the inertial Solitary Kelvin mode. It is clear how the wave crest precedes the whole body of the wave in the same direction of the system propagation, which is counterclockwise. Russell (1844) stated that the wave breaks when its amplitude equals the depth of water under it, which is the case in Figure (11), $\bar{h} = 9.33$cm, $A = 9.17$cm. In the transverse direction on the other hand, it was noticed that the wave has exponential decay form away from the wall, without curving backward in that direction. The radial wavenumber that is used in the experiment is given by:

$$M = \frac{\pi}{1-\beta}, \quad (39)$$

where $\beta = \frac{r_{min}}{r_{max}}$ the ratio between the inner and the outer radiuses as in table (2). Simple fit was used assuming exponential form as follows:

$$A_r = A_m e^{(Mr-\beta_1)} + B, \quad (40)$$

where $B$ is an arbitrary constant that suits the fit, and it is different in each experiment, $A_m$ the wave amplitude, $r$ the radial distance between the inner and the outer cylinders, $\beta_1$ is dimensionless Rossby Radius, which is given as:

$$\beta_1 = C/\omega, \quad (41)$$

where $C$ the shallow water velocity: $C = \sqrt{g\bar{h}}$, and $\omega$ the frequency of oscillation. This exponential fit for the radial amplitude decay is presented in Figure (12) in comparison with the experimental results.

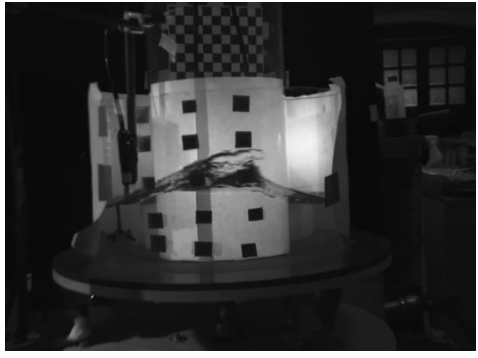

**Figure (11).** Breaking of the solitary fundamental Kelvin mode, the crest precedes the wave body, volume = 10000ml, $\tau = 0.02333$, $\Omega = 6.25$ rad/sec

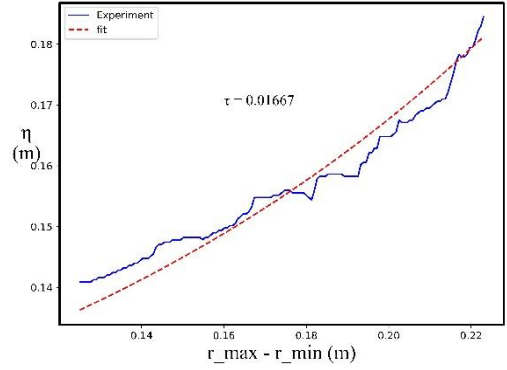

**Figure (12).** Radial exponential decay of solitary wave amplitude, volume 10000ml, $\Omega = 6.84$rad/sec dashed line is the exponential fit.

To estimate the rotation and viscous effects, a computation of Reynolds, Ekman, and Rossby numbers was carried out, for each Solitary Kelvin wave mentioned above, at different volumes of water. Reynolds number determines the type of the flow, like laminar and turbulent. In open channels it can be given in terms of its characteristic length, the hydraulic radius, in the system under study it is assumed as:

$$R_e = \frac{\Omega R_h^2}{\nu}, \qquad (42)$$

where $\nu$ is the kinematic viscosity which is has a constant value assumed during the experiment $1.03 \times 10^{-6}$, $R_h = A/P_w$ the hydraulic radius, $P_w$ the wetted perimeter: $P_w = r + 2\bar{h}$, as the cross-section for the channel has rectangular form, with $r = r_{max} - r_{min}$ the channel width and $\bar{h}$ is the average depth, $\Omega$ the rotation speed. It is known that Reynolds in open channels has different ratios in comparing the turbulent and laminar effects. If $R_e \leq 500$ the flow is laminar, and $R_e \geq 1000$ the flow is turbulent. In this system, the computations showed that for resonant conditions that agree with Kelvin Solitary wave, the flow is always turbulent. To measure the rotational effect Rossby number is computed, which is a dimensionless number that represents the ratio between inertial and Coriolis forces, it is assumed in this system as in (43):

$$R_o = \frac{\Omega_1}{2(\Omega_1 + \Omega)} \leq 1, \qquad (43)$$

where $\Omega_1 = \tau\Omega$ precession rate. The computations of Rossby number showed that for all the above-mentioned cases of tilt is constant and equal to 0.492, which shows moderate rotation effect on wave motion. Ekman number represents the balance between the small viscous forces and the small Coriolis ones, and it plays an important role in determining the boundary layer effect on the solid walls. It is assumed to be the inverse of Reynolds number as it is clear in (44), to ensure that the inviscid effect is satisfied. The computations of this number showed that it is of order $O(10^{-4})$:

$$Ek = \frac{\nu}{\Omega R_h^2}. \tag{44}$$

As we deal with open channel, computations for Froude number using (45) showed that the flow all the time supercritical regardless the volume of water in the channel:

$$F_r = \frac{r_{max}\Omega}{\sqrt{g\bar{h}}}. \tag{45}$$

The velocity of the wave was measured by taking the difference in time between two different pictures, from which the difference between the wave crest in each one is approximately $\lambda = 2\pi r_{max}$, which is the crossed distance per time. Amaouche *et al.* (2013) proposed that the velocity in their tank theoretically was given by:

$$c_{th} = \sqrt{g\bar{h}}\left(1 + \frac{1}{2}\frac{A}{\bar{h}}\right). \tag{46}$$

In comparison with this work, it was found high match with experiment as shown in table (3):

**Table (3).** Comparison between the theoretical wave velocity and one computed experimentally.

| Volume | $c_{th}$ | $c_{exp}$ | Err % |
|---|---|---|---|
| 10000 | 1.528 | 1.547 | 0.0124 |
| 12000 | 1.463 | 1.342 | 0.0827 |
| 14000 | 1.575 | 1.521 | 0.0343 |

# 4. Conclusion.

In this paper, a new KdV model equation was derived for the case of the flow in an open cylindrical channel under precession conditions. The derivation starts from assuming the inviscid irrotational conditions in two-dimensions of the problem, as the radial part is short and barely affects the dominant azimuthal flow direction. The derivation first was to the forced Bernoulli equation, then it followed the classical perturbation method to derive the final version of this eqaution. The equation has a forcing term, that makes the equation is not solvable directly, and a simple Finite Difference Scheme is used to solve it. However, as noticed from the experiment that this term is small and can be neglected from the equation and kind of analytical solution was derived, accordingly. Both the numerical and the analytical one were compared with the observed wave in the experiment, with very good agreement. It is worth to mention that during the experiment an assumption was made for Rossby radius that takes the local frequency, not its double value. The radial variation of the waves observed in this system has exponential decay, and the wave crest was not curved backward. It was noticed that the effect of rotation in this case is severe where once the wave amplitude is bigger or equal to the average depth of water it either breakes or disperses, in waxing and wanning mode from which no recovery to its original form, this is called the resonance collapse for the case of forced oscillations. Other thing is worth to mention that the breaking effect when starting from rest conditions is similar to the normal rotation effect in gas dynamics (Ursell 1953) where the observed waves have crests proceed the whole body of the wave. Although the velocity of the wave is taken the solid-body rotation one, another one that was proposed by Amaouch *et al.*

(2013) based on the long wave one was used. It was in a perfect match with the wave velocity noticed in the experiment.

Acknowledgements

The author is thankful for Professor Szabó Gábor for suggesting this beautiful interesting idea, the author is also thankful for Professor Zsolt Kohári for writing her the curve tracking code, and for Professor Yiannis Hadjimichael for his help on the finite difference method, and Mariann Szilágyi for her help in tracking the wave experimentally. The author is thankful for the unknown referees for the invaluable suggested references and comments on the early version of this manuscript.

Funding

The work was supported by the **Stipendium Hungaricum Scholarship** under contract **no. SHE-15651-001/2017** to *Hajar ALshoufi*, Budapest University for Economics and Technology, Budapest, Hungary.

Data Availability Statement

The data that support the findings of this study are available from the corresponding author upon reasonable request.

Formal Contribution

The present work in this paper is totally carried out by the author: *Hajar Alshoufi*. Where she wrote the manuscript, derived the equations, carried out the experiments, solved the problem numerically, and compared the results with the experiments.